\documentclass[11pt]{article}
\usepackage{blois,epsfig}

\def\be{\begin{equation}}
\def\ee{\end{equation}}
\def\bea{\begin{eqnarray}}
\def\eea{\end{eqnarray}}

\begin{document}

\vspace*{1 cm}
\title{The 1953 Cosmic Ray Conference at Bagneres de Bigorre: \\the Birth of Sub Atomic Physics }

\author{ James W. Cronin } 

\address{Department of Astronomy and Astrophysics\\
Enrico Fermi Institute, University of Chicago\\
5640 South Ellis Ave. , Chicago, IL 60637, USA}

\maketitle\abstracts{The cosmic ray conference at Bagn\`{e}res de Bigorre in July, 1953 organized by Patrick Blackett and Louis Leprince-Ringuet was a seminal one. It marked the beginning of sub atomic physics and its shift from cosmic ray research to research at the new high energy accelerators. The knowledge of the heavy unstable particles found in the cosmic rays was essentially correct in fact and interpretation and defined the experiments that needed to be carried out with the new accelerators.
A large fraction of the physicists who had been using cosmic rays for their research moved to the accelerators.
This conference can be placed in importance in the same category as two other famous conferences, the Solvay congress of 1927 and the Shelter Island Conference of 1948.}
\section{Introduction}

In January 2010 the Large Hadron Collider (LHC) at CERN began to produce proton-proton collisions at a center of mass energy of 7 TeV.
This machine is the most complex and most costly  of a long series of accelerators. Results of these accelerators have led to a detailed understanding of how the subatomic world works. However fundamental questions remain and it is hoped that the LHC when it achieves its full energy (14 Tev) and intensity will answer some of these fundamental questions.  The accelerators have been the mainstay of subatomic
physics (or high energy physics or elementary particle physics) since 1953 when the 3 GeV Brookhaven Cosmotron began artificially
producing the heavy unstable particles found in the cosmic rays. In July, 1953, a  conference was held in the French Pyrenees town of 
Bagn\`{e}res de Bigorre which was devoted entirely to the production and decay properties of the cosmic ray discoveries. With perhaps one exception, concerning the production of $\Lambda$ hyperons, all
the conclusions concerning the unstable particles were correct. The cosmic ray results defined the early experiments to be carried out at the
accelerators. The properties of the production and decay of these particles were sufficiently known so that Abraham Pais\cite{Pais} and 
Murray Gell-Mann\cite{Gell-Mann} and Kazuhico Nishijima\cite{Nishijima} could predict their production in pairs and the strangeness scheme which defined permitted and forbidden modes of production and decay. 
This conference marked the boundary in time when the field of subatomic physics passed from cosmic ray research to
the accelerators. This shift was explicitly recognized at Bagn\`{e}res de Bigorre. 
This article tells the story of this remarkable conference.

\section{Planning the conference}

In early March 1952 Louis Leprince-Ringuet received a pleading letter from
Patrick Blackett. In 1950   Blackett had been chosen president of the Cosmic Ray Commission of the International Union of Pure and Applied 
Physics (IUPAP) and Leprince-Ringuet the secretary. Blackett writes:   \cite{B1}\\

 {\em I wonder if you have been able to find out anything about any plans that may have been made? Who has the minute book? Surely we should be discussing future plans? I do not even know the members of the commission!}\\

In 1952 the members of the commission in addition to Blackett (United Kingdom) and Leprince-Ringuet (France) were Carl. D. Anderson (United States), Gilberto. Bernardini (Italy), Homi J. Bhabba (India), and Manuel Sandoval Vallarta (Mexico). The Commission had previously organized conferences
at Cracow(1947), Como(1949), and Bombay (1950). There are no known  published proceedings for these conferences. At Bombay the Commission decided that the next conference should be held in 1953 and  Blackett expressed alarm that nothing had been done!\\

In the following months Blackett and Leprince-Ringuet began the organization for the 1953 conference.
It was decided to hold the conference in July 1953 at Bagn\`{e}res de Bigorre, a town close to the high altitude French Observatory at Pic du Midi, located in the Pyrenees. The Pic du Midi Observatory was managed by the University of Toulouse and directed by Professor Jean  Rosch. The University of Toulouse became co-sponsor of the conference with IUPAP. \\

 On May 19, 1952 a letter was sent from Blackett and Leprince-Ringuet \cite{LR1} to the above mentioned commissioners outlining the general details of the conference and the subjects to be covered. The important passage in the letter is quoted below:\\
 
 {\em We feel that the whole field of cosmic rays is now too wide to be dealt with in a single conference, and it would be therefore better to limit the subject to certain lines of special contemporary interest. The general subject would  be `Interactions at Ultra Relativistic Energies' including the creation of V, K, $\tau$, and
 $\xi$ particles, cosmic ray phenomena underground, the primary spectrum, and any recent results related to the general definition given above. 
 You will notice that we have excluded  in the main the geophysical aspects of cosmic rays, since we feel that these are of sufficient importance to justify being made the subject of another conference, perhaps in 1955, along with a discussion of the theories of cosmic rays.}\\

Leprince-Ringuet always wrote in French regardless of the nationality of the recipient, the sole exception being the joint letter drafted by Blackett to the commissioners. All letters to Blackett were in French. All from Blackett were in English.\\

The first reply was from Bhabba on May 26 who stated that he preferred an earlier date to better conform with the academic schedules in India.\cite{Bh1}\\

On May 20 a letter from Blackett to Leprince-Ringuet \cite{B2} reported that a letter from Pierre Fleury,  Secretary General of IUPAP, said that very few funds would be available and suggested that the conference be delayed until 1954. Blackett took exception to this, writing:\\

 {\em I hardly think we should wait until then but should try to get on with less money.  I don't see any reason really why we should not have a conference because we cannot pay the American's fares over here!  Quite a number of Europeans will be able to come anyway whether they are helped or not. So I think the best policy is to press for the maximum amount we can get from UNESCO but to go ahead with the plans anyway.}\\
 
One can sense the disdain in the reference concerning ``the Americans". At this point in his life Blackett had made many contributions to the British war effort essentially having invented Operations Research which among many applications led to the elimination of the submarine menace by 1943.  After the war he returned to research  at Manchester but he was still much involved in the scientific issues concerning nuclear weapons and had a very negative opinion of the US buildup of offensive weapons for nuclear war. He was, for this reason, a controversial figure in his own country. At this time Manchester was very active in cosmic ray research but Blackett was less directly involved.   \\

On May 26 Leprince-Ringuet replied to Blackett \cite{LR2} agreeing that the conference should proceed despite the funding uncertainty. He offers to look into the budgetary situation. Following his promise to look into the finances in his May 26 letter to Blackett,  Leprince-Ringuet writes to Fleury on June 5 \cite{LR3}. In the letter he states that he and Blackett are determined to have the conference in July 1953. He states that the other 4 members of the Commission have been informed of the proposal (letter of May 19) although only one response had been received at the time. Leprince requests that Fleury
inform the President of IUPAP , Professor Neville Mott, that it is the intention of the cosmic ray comission to hold the conference and that 
if either the President or other members of the IUPAP executive committee have any objections, please inform him.\\

On June 19 Leprince receives a very encouraging letter from Fleury \cite{Fl1}. It seems that Leprince had much more influence on the Frenchman
Fleury than Blackett. Fleury's letter reads in part.\\

{\em    The proposition made by Mr. Blackett and yourself to organize at Bagn\`{e}res de Bigorres in July 1953 a reunion of the IUPAP Commission on Cosmic Rays and a conference on certain problems on cosmic rays has been examined by our executive committee
on the 6 and 7 of June.\\

The committee has decided to offer the support of IUPAP for these reunions and to request that UNESCO contribute to the costs the
following support.:\\

$\$$1500 for the commission (travel and subsistence)

$\$$1500 for the conference (travel and subsistence)

$\$$300 for the publication of the proceedings \\}

It is curious that the request to UNESCO included an amount to support the attendance of the six members of the Cosmic Ray commission equal to its support for  conference proper! The UNESCO funds were to be used for expenses incurred outside of France. \\

The letter went on to say that the response of UNESCO could not be expected before December 1952, however one could develop a detailed program for the conference and develop a list of potential participants which had to be approved by Professor Mott! The letter stressed
that no financial commitment could be made in advance of the receipt of the UNESCO funds. The letter went further stating that each participant be prepared to pay their travel and subsistence in the event that the available support fell  short of the request.\\

The second commissioner to reply to the May 19 letter was Vallarta \cite{V1}, who wrote on June 19 that the program outlined in the May 19 letter was fine but should include:\\

{\em  ....some reports  on the geomagnetic aspects of cosmic rays, in particular the question of the albedo,
for which recently a lot of progress has been made both in theory as well as experiment. It seems to me that such a report can evoke a certain interest. }\\

Incidentally Vallarta apologizes for the bad quality of his French, which in fact was perfect. He mentions learning French from his 
grandmother when he was a small child.\\

On July 7 in a letter to Blackett \cite{LR4} , Leprince-Ringuet  reports on the meeting of IUPAP.  He asks Blackett to personally urge Professor Auger, scientific director for UNESCO, to not reduce the funding for the conference from what had informally been promised.
Then Leprince reports that of the commissioners  two of the four had not responded to the their joint letter of May 19. He suggests that the failure to respond reflects agreement with the proposed limited scope of the conference. As for Vallarta, Leprince-Ringuet writes:\\

{\em On the other hand Vallarta proposes, as one could predict, an extension of the conference to include geomagnetic aspects; we have suggested in order to not overload our planned program that one have another congress to cover the geomagnetic phenomena. Are you still of this opinion?\\

 If you do not think the responses of the Commission sufficiently favorable we can write a letter to Anderson and Bernardini reminding them of our letter of May 19. Otherwise we can right away prepare a detailed program and a list of people which we can submit to the Commission. If you agree we can prepare this first program roughly with a list of
 people to invite in each of the categories }\\ 
 
On November 11 Leprince received a letter from Bernard Peters \cite{P1} at the TATA Institute (India) which referred to a letter 
Peters received from Leprince-Ringuet dated October 20. (This letter was not found in the  archives).
The presumed letter outlines to a broader group of cosmic ray researchers the restricted scope of the conference.
Peters suggests that the conference be extended to production processes of the new particles, and as well, that ideas for new experimental techniques be discussed. He suggests that reports on early accelerator work be included. (The
 3 Gev Cosmotron at Brookhaven National Laboratory began running in 1952).  As will become apparent, if this suggestion had been followed more vigorously, one of the conclusions of the conference would have been even closer to the truth.  Most importantly he suggests that Russian scientists be invited well in advance. \\
 
 On November 20 a second letter was sent to the commission \cite{LR5}. By this time the dates for the conference had been chosen ( 6-12 July, 1953).  A choice between two programs was presented, either a discussion only of the
 properties of the new particles or a more extended program where the interactions at high energies, underground interactions and stars in emulsion would be included. The length of the discussion of the new unstable particles would have to be reduced by
 two days to accommodate  the extended program, and was therefore not preferred by Leprince-Ringuet and Blackett. Also
 a suggestion by Daudin  that a restricted group studying geomagnetic effects would discuss in a parallel session the standardization  counting equipment used in these studies was included. \\
 
 Professor Rosch had been working hard on the organization and the funding of the conference.  A letter dated the 24th of November 1952
 to the Rector of Toulouse University \cite{Ro1} reviewed the work that had been done. This included contact with IUPAP, a meeting with the director of higher education in France, a full budget of 5 million francs ( about $\$$14,000) raised from UNESCO, the department of higher education, the
 department of cultural affairs, the local departments of Haute Garonne and Haute Pyrenees, and the University of Toulouse. He also established a local committee chaired by himself to make all logistical arrangements leaving Leprince-Ringuet and Blackett to worry only about the science program.  \\

\begin{figure}[!ht]
\begin{center}
\includegraphics  [width=1.0\textwidth] {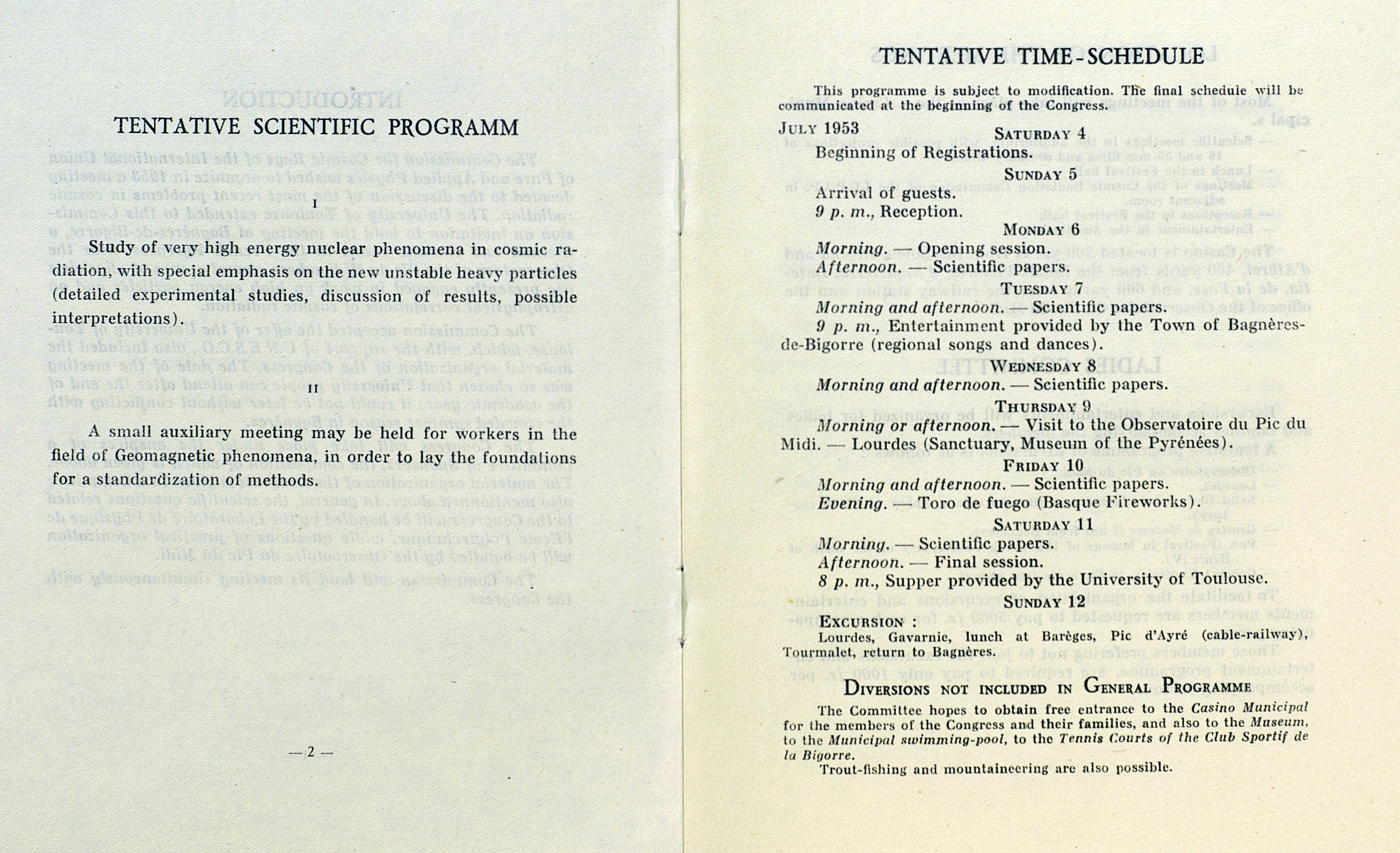}
\end{center}
\caption{\it Two pages  from the conference announcement , March 10, 1953.}
\label{fig:Bagn1}
\end{figure}

 Letters from Bhabba \cite{Bh2} on December 1, 1952  and Vallarta \cite{V2} on December 9, 1952  favored the extended conference.
 But ultimately the conference was restricted to the properties of the new particles.  This program was even more restrictive than proposed in the original letter of May 19, 1952 to the Commissioners. This decision was taken  despite the fact that Vallarta and Bhabba continued to argue for the larger conference. Of the six Commissioners Anderson, Blackett and Leprince-Ringuet prevailed.
However Daudin's specialized meeting was supported.  By this time Bernardini could no longer be considered a "cosmic ray man" He
had taken a position at Illinois and was working on photo-meson production at the Illinois betatron. On June 10, 1953 he wrote an apologetic letter \cite{Br1} to Leprince-Ringuet questioning the appropriateness of his membership on the cosmic ray comission of
 IUPAP.\\

In early December UNESCO gave a positive response for the support \cite{Fl2}. They allocated $\$$1650 for conference support, $\$$800 for Commission support and $\$$300 support for publication. The total sum was less than requested but the balance between the conference support and the commission support was much more reasonable. One can speculate that this alteration was the result of phone calls between Leprince-Ringuet and  Pierre Auger.  \\

 On March 10 1953 an announcement of the conference  was mailed out with the logistical details.  The purpose of the conference and the tentative scientific program is shown in Figure 1.\\

{\em Study of very high energy nuclear phenomena in cosmic radiation, with special emphasis on the new unstable heavy particles (detailed experimental studies, discussion of results, possible interpretations)\\
 
A small auxiliary meeting may be held for workers in the field of geomagnetic phenomena, in order to lay the foundations for a standardization of methods.\\}
 
 The program was to consist of three days (morning and afternoon) of scientific presentations, mostly experimental. A crucial addition was
 to arrange three subsequent sessions each led by an outstanding physicist to merge if possible the disparate reports of the first three days into a coherent picture.  It was this format that produced a conference remembered by many participants as the best conference 
 they had ever attended. The persistence of Leprince-Ringuet was well rewarded.  \\

 \section{Correspondence with delegates to the conference}
 Following Peter's suggestion Leprince wrote on March 12 the president of the Soviet Academy of Sciences inviting the representatives of the Soviet cosmic ray physicists to attend the conference. \cite{LR6} He writes in part:\\
 
 {\em We are eager to have the participation of all the countries active in the studies of cosmic rays and especially in the analysis of the properties of the new heavy unstable particles discovered recently in the nuclear interactions of very high energy cosmic rays.\\
 
 Excellent studies have been made in your country in this domain. We understand the difficulty of making such a long trip so we ask you to
 designate three or four representatives whom you deem the most qualified. ...........\\} 
 
 On June 22, two weeks before the beginning of the conference, Leprince received an answer from the scientific attache at the Soviet embassy in Paris.\cite{So1}\\
 
 {\em I have the honor to forward the letter from the President of the Academy of Sciences of the USSR, Mr. Heseianov, in which the Academy of Sciences of the USSR thanks the Academy of Sciences of Paris for its friendly invitation to take part in the International Congress for cosmic rays.\\
 
 Unfortunately the Soviet scientists who work in this area will be occupied in an expedition this summer and cannot take part in the 
 congress.\\}  

This was the expected response yet disappointing as many cosmic ray physicists were eager to resume scientific contacts with the Soviet Union. In 1955 a number of western particle physicists were invited to visit the Soviet Union. They were shocked to find that the the Soviets had built a 600 MeV synchrocyclotron and a 10 GeV synchrotron. Contacts were gradually resumed but until 1989 Soviet scientists had great difficulty in traveling and were always accompanied by colleagues who were there not for science but for a watchful eye.\\ 
 
 Leprince-Ringuet was relentless in his determination to have  the leaders in the field of cosmic rays attend the conference.  The conference was structured so that the first 6 half days (Monday-Wednesday) were devoted to short reports on the experimental results. Thursday was devoted to excursions. Friday morning, as a concession to Bhabba, there was to be a session on high energy interactions. On the following three half days there was to be recapitulation of photographic emulsion results led by C. F. Powell, a recapitulation of cloud chamber results led by W. B. Fretter, and finally a synthesizing of all the results by Rossi. Leprince had these sessions well in mind when he wrote these discussion leaders  in advance of the formal announcement of the conference with the mailing of March 10, 1953.\\
 
 On December 10, 1952, Leprince-Ringuet wrote Powell:\cite{LR7}\\
 
{\em Before even sending out the invitations, I am writing this letter to tell you that we most eagerly want you to participate in this congress:
if you are not there one of the most interesting aims of the conference will be lacking, that is to say an in depth discussion  led by the principal specialists.}\\

In early 1953 Leprince wrote W. B. Fretter:\cite{LR8}\\

{\em  We just received a word from Blackett who is very happy to have you lead the general  discussion on the results from the Wilson chambers at the congress de Bagn\`{e}res}\\

On March 10, 1953 Leprince wrote Rossi:\cite{LR9}\\

{\em  ....... It goes without saying we count in a very strong fashion on your participation. You are one of three or four physicists without whom the conference will not succeed. ..............  If you look at the program you will see that the last half day is dedicated to a synthesis of all the 
results. We count on you to lead this last session.}\\

Leprince saw the conference as a way to demonstrate the renewed vitality of science in France following the war.
He wrote personal invitations  to a number of prominent scientists including Fermi, Bethe, Goldhaber, Heisenberg, and Oppenheimer. None
of these attended. What was of interest was that a number of physicists  had heard about the conference via the grapevine and wrote to request invitations either for themselves or colleagues. Among these were Abraham Pais and Richard Dalitz who were to make enormous contributions to particle physics based on the data that was collected and systemized during the conference.\\

For example Dalitz wrote\cite{Dalitz}:\\

{\em ....... I wish to contribute to the second part of the conference a theoretical paper entitled " The modes of decay of the $\tau$ meson" to occupy 15 minutes}. \\

There were also amusing letters.\\

A Minnesota professor, addressing Leprince as ``Petit Prince'', wrote:\cite{Ney}\\

{\em I would like very much to attend the conference in the Pyrenees in July. .............It would be very good if I could locate some little French girl to teach me the language before I come over. I am looking forward to seeing your ``charming'' scanners.}\\

\section{The Conference}

The Proceedings of the Bagn\`{e}res de Bigorre conference are difficult to find\cite{Proceedings}. Shown in Fig 2 is a photo of the cover page of a copy of the  proceedings found in the Leprince-Ringuet archives. There was no formal publisher and the proceedings only exist in mimeograph copies presumably sent to each attendee.  But they cover the conference in detail. Particularly important is the reproduction of the discussions which fill about 20$\%$ of the 300 page proceedings. The secretaries who recorded and supervised the typing provide a
disclaimer:\\

\begin{figure}[!ht]
\begin{center}
\includegraphics  [width=1.0\textwidth] {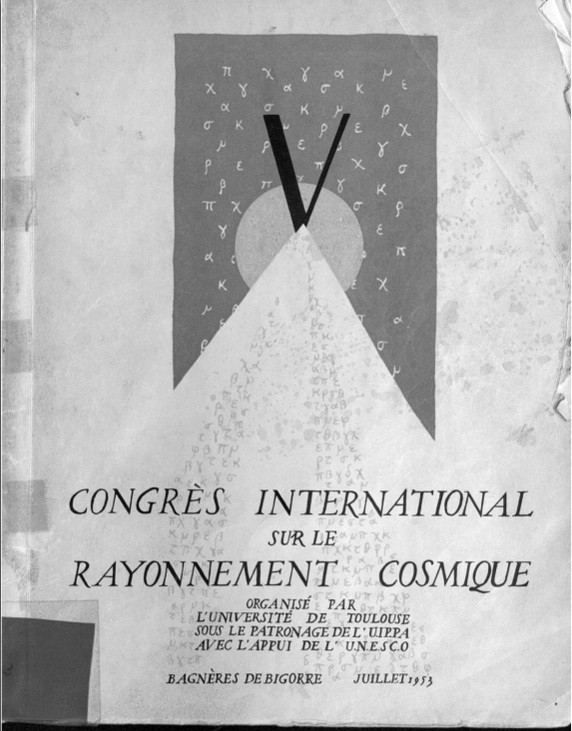}
\end{center}
\caption{\it Cover page of the Bagn\`{e}res de Bigorre Proceedings}
\label{fig:Bagn2}
\end{figure}

  {\em The secretaries have attempted to write these proceedings as faithfully as possible, no proofs were sent for correction to the members of the conference. Any error which may appear is the sole responsibilities of the secretaries.}\\
  
Their work was heroic but their names remain anonymous. There was no indication in the archives as to their identity.\\
   
In the preface of the proceedings was written:\\

{\em The particles described in this conference are not entirely fictitious and every analogy with the particles really existing in nature is not purely coincidental.} \\

A photo of the attendees at the conference is shown in Fig 3.  Leprince-Ringuet and Blackett, in dark suits, stand in the middle of the front row. Rossi in a light double breasted suit stands to their right  and a bit behind.

\begin{figure}[!ht]
\begin{center}
\includegraphics  [width=1.0\textwidth] {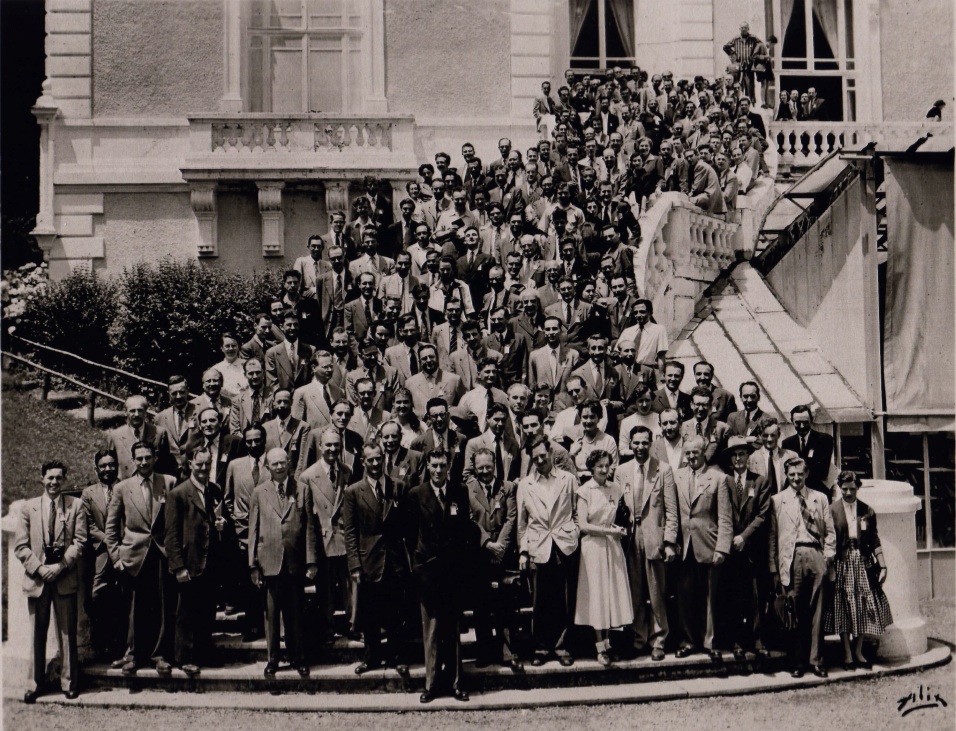}
\end{center}
\caption{\it Official photo of the conference}
\label{fig:Bagn3}
\end{figure}

Blackett, in his opening remarks says:\\

{\em The organizers of this Conference gave much thought to what form it might usefully take; finally it was decided to limit the scope of the Conference to one part only of the wide field of cosmic-rays, that of the nature and behavior of the unstable elementary particles occurring in the cosmic radiation . We chose this subject because of the exciting new knowledge which is now forthcoming from so many laboratories.
We excluded other aspects of the great subject of cosmic rays, the geophysical aspect, primary spectrum, extensive air showers, time variations and theories of origin in order to prevent the Conference from becoming too large in size and too long in duration}\\

The conference proceeded exactly as planned. During the first three days the experimental data on the heavy mesons (K particles)
and particles more massive than nucleons (hyperons) were presented in detail by the experimenters. Also the experimental techniques were discussed along with their limitations. There are very few physicists remaining who were at the conference and have a clear recollection of the details. Among these is Donald Perkins\cite{Perkins} who had clear reminiscences of the conference. He recalled two outstanding contributions to the conference. One was the work of Robert Thompson\cite{Thompson} of Indiana University who made precision cloud chamber measurements  of the decay
$\theta$ $\rightarrow$ $\pi^{+}$ + $\pi^{-}$. The other was Richard Dalitz\cite{Dalitz} with the analysis of the spin/parity of  $\tau$ $\rightarrow$ 3$\pi$.\\

On Thursday July 9, while the bulk of the conferees went on excursions or played at the casino, Powell, Fretter, and Rossi struggled to
pull all the contributions together.  On Friday afternoon July 10, Powell led a discussion intended to summarize all the information on the heavy mesons.  Most of these data came from emulsion experiments for which Powell was an expert. On Saturday morning Fretter led the discussion concerning the cloud chamber data. In both of these sessions all of the principal experimenters were present. Extensive discussion covered the disagreements and ambiguities. As we know now  the nature of the unstable particles and their properties, it is difficult to imagine the confusion produced by a myriad of results on new phenomena containing the ``noise" of experimental errors and incorrect
results. It is remarkable that so much clarity was brought forth at Bagn\`{e}res de Bigorre with the remaining uncertainties quite well defined.
Much credit for this must go to Bruno Rossi who made the final summary of the conference.\\ 

Rossi was determined to simplify the nomenclature for the heavy mesons and the unstable particles heavier than protons. The particles were denoted by Roman characters. The decay modes were denoted by Greek letters. In Figure 2 is shown Rossi's classification as printed in the proceedings.  In advance of the conference Rossi was planning to introduce this systematic nomenclature for the unstable particles. Rossi's thoughts on the subject were expressed in a letter from Herb Bridge to Bernard Gregory\cite{Bridge} which states in part:\\

{\em Bruno would like to propose a scheme for classification of the new unstable particles. He wants to use Greek letters to a postulated decay scheme. The phenomenological definitions (which use Latin letters) can then be compared quickly with a given decay scheme
by mentioning the appropriate Greek letter. I enclose the proposal.\\

It is easy to see that getting agreement on this at the conference might start an argument which could waste a lot of time.  Bruno is 
particularly anxious to avoid any result of this nature and wants to circulate the proposal ahead of time to the interested people. He
feels that in this way one might get a definite and useful agreement at the conference. }\\

\begin{figure}[!ht]
\begin{center}
\includegraphics  [width=1.0\textwidth] {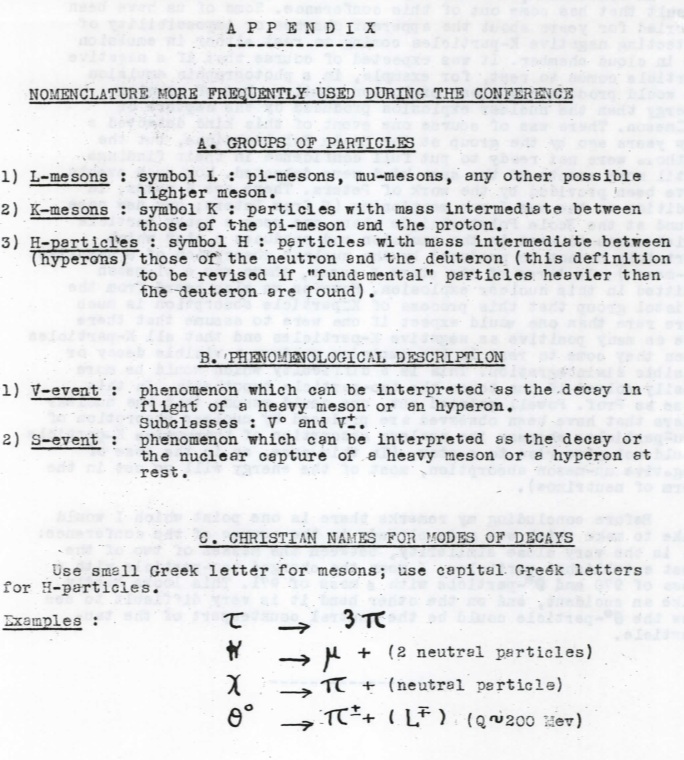}
\end{center}
\caption{\it Rossi's nomenclature for the new unstable particles}
\label{fig:Bagn4}
\end{figure}

Rossi organized his summary into two parts. The first discussed the unstable particles heavier than the proton. The first topic was the
V$^{0}_{1}$ now known as the $\Lambda^0$ hyperon. It was well established with a decay V$^0_1$ $\rightarrow$ p + $\pi^{-}$ with a mean life of $\sim$ 3 x 10$^{10}$ sec and a Q-value of 37 MeV.  The issue was the mode of production.  Marcel Schein  presented evidence that the V$^{0}_{1}$ could be produced by 227 MeV $\pi^{-}$. Rossi concluded ``So it is quite safe to assume today that the V$^0_1$-particle can be considered as an excited state of the neutron; this simply means that the V$^0_1$-particle is created out of an existing nucleon.''  Rossi went on to say `` The other question is whether the { \underline {V$^0_1$ particles are formed individually or in pairs.}}'' and added ``........the question is to know whether they can be made singly at all, because the theorists would like to think in order to account for the comparatively long mean-life of the V$^0_1$ particle that they can only be produced in pairs. As far as I can see the only possibility for such pair production would be:  $\pi^-$ + p +n $\rightarrow$ V$^0_1$ + V$^0_1$ '' .  \cite{Schein}
This conclusion of Rossi, a thoughtful and critical physicist, without any expression of skepticism is quite surprising. How could a low energy pion excite simultaneously two nucleons? Perhaps it was his subtle way of stating ``reducto ad absurdum''.\\  

Missing from the discussion was the recent detection of a V$^0_1$ in a diffusion cloud chamber at the newly operating Cosmotron at Brookhaven National Laboratory. In his reply to the invitation of Leprince-Ringuet, Maurice Goldhaber had written on April 3 that an invitation should be sent to R. P. Shutt to
describe the first accelerator production of a V$_1^0$. There is no evidence in the archives that Shutt had been invited. Had Shutt or a Brookhaven colleague attended, the conclusions about V$_1^0$ production might have been more
tentative. Pais \cite{Pais} had written a paper in which V$_1^0$ would be produced with another unstable particle. Despite the fact 
that Pais was listed as an attendee at the conference he in fact did not attend. His non-attendance was confirmed in his autobiography \cite{Pais1}.\\

Rossi continued with his summary.  There was evidence presented for a charged particle heavier than a proton the V$_1^+$. Clear evidence was presented for a decay V$_1^+$  $\rightarrow$ n + $\pi^{+}$ with a Q value of about 133 MeV. Less sure ``but still a possibility'' were two cases of a decay of  V$_1^+$  $\rightarrow$ p + $\pi^0$. The existence of this particle was ultimately established as the $\Sigma^+$. 
This positive charged hyperon was to be found in Gell-Mann's \cite{Gell-Mann} organization of the new unstable particles.

Rossi then reports: `` One of the great surprises of this conference has been the additional evidence presented by Leighton for the existence of the so called {\underline {cascade decays}} revealed by the decay V$^{*-}$ $\rightarrow$ V$^1_0$ + $\pi^-$ . The
cascade particles were soon named $\Xi $ particles.
Rossi's final remark about unstable particles heavier than a nucleon is  that a V$_1^0$ can replace a neutron in a nucleus; this phenomenon  led to the study of hyperfragments in photographic emulsions.\\

Rossi then turned to the mesons that were heavier than the $\pi$ and $\mu$. One of the most important contributions to the conference was given by Robert Thompson of Indiana University. He presented measurements of the $\theta^0$ $\rightarrow$ $\pi^+$ + $\pi^-$ observed in a precision cloud chamber. There was additional evidence presented at the conference that the decay particles had nuclear interactions, confirming that they were $\pi$ mesons and not $\mu$ mesons. The $\theta^0$ was
well established and had a mass of 971 $\pm$ 10 electron masses (m$_e$).\\  

The decay $\tau^+$ $\rightarrow$ $\pi^+$ + $\pi^+$ + $\pi^-$ was also well established with measurements in nuclear emulsions. Its  mass was found to be 970  $\pm$ 5 m$_e$.\\

Much evidence was presented for other decay modes of the heavy mesons and the possibility that there were heavy mesons with masses
around 1200 m$_e$ as well as 1000 m$_e$. The situation was confusing and led to much discussion in the summary sessions led
by Powell and Fretter.  Rossi bravely tried to bring some order to the evidence. Important in this effort was the introduction of a nomenclature
reproduced in Figure 4.  All heavy mesons were placed in a general category K. Only the cases discussed above were considered established. However Rossi took the point of view " that two particles are equal until proved different". As time passed Rossi's hunch
held true and all the heavy mesons discussed turned out to be either neutral or charged K mesons. There was another problem in that there
appeared to be more K$^+$ than K$^-$ found in the data. This prompted the prophetic comment by Robert Sard : ``Maybe we can assume that 
K$^+$ and K$^-$ particles are not produced in equal numbers.'' This production asymmetry would be supported by the Gell-Mann Nishijima
strangeness scheme \cite{Gell-Mann} submitted in August of 1953. (Gell-Mann's strangeness scheme is out lined in an appendix) \\

At the end of his summary Rossi adds:\\

{\em Before concluding my remarks there is one point I would like to make which was made already in the course of this conference: it is the very close similarity , between the masses of two of the best established particles, I mean the charged tau-particle with a mass of 970 and the 
$\theta^0$ with a mass of 971. This looks hardly like an accident, and on the other hand it is very difficult to see how the  $\theta^0$-particle
could be the neutral counterpart of the tau-particle.\\}
 
Concluding remarks were made by Blackett and Leprince-Ringuet.\\

The IUPAP Cosmic Ray commission met during the meeting. Blackett reviewed some of the recommendations which principally
said that the next Cosmic Ray conference in 1955 would concentrate on Geophysical and Astrophysical aspect of cosmic rays.
All subsequent conferences have pursued the same theme. Concentration on particle physics shifted to the ``Rochester Conferences''
which evolved into the International Conferences on High Energy Physics. Turning to the present conference Blackett remarks:\\

{\em I think it has been in many respects the best (conference) I have ever attended, especially for the sustained interest, the high level of discussion and the friendly but sharp, critical attitudes of the participants. It has also been remarkable for the high level of attendence.\\

I think this conference will have a very good effect on experimental workers. I feel that all young physicists  who have attended this conference (and not a few of the older ones) will be a little more careful when publishing their next paper to ensure that they have not overlooked any systematic error or underestimated their random error.\\

So, when we meet again we may hope to have so improved our accuracy that many of the controversies of this week will be quite settled.
However if the history of scientific discovery is any guide, the same increases of accuracy which will serve to settle or present controversies will equally, surely, itself bring to birth new controversies by leading to some new discoveries. }\\ 

Leprince-Ringuet's concluding remarks were extensive, commenting on the conference and looking to the future with remarkable clarity.
Here follows (in rough translation) some of his remarks.:\\

{ \em We have hoped, as Blackett told you at the beginning of this busy week, that
this conference would have a narrow focus. Indeed it has been very narrowly 
focused and also very serious. This was our goal: a series of presentations and
discussions which could truly advance our knowledge of the fundamental 
particles observed in the cosmic rays. I have been struck by the beautiful
scientific atmosphere, the precision, the rigor, and also by the friendly 
character and fellowship of the discussions, particularly during the 
final sessions of the conference. A major scientific clarity has blown away the fog of confusion
and all of us are extremely happy.\\

I must confess that Gregory and I had some anxiety a few months ago when we 
left empty, with no program whatsoever, the last three sessons.They 
were intended only for discussions and conclusions. We worried that these 
sessions would be poorly attended, incompletely filled, and interest in 
them would fade before the end. But not at all! We had to insist that the 
sessions began and ended on time! We have obtained, after the masterful 
presentations of the leaders of the last three sessions, Powell, Fretter, 
and Rossi, the knowledge of not only a large amount of reliable data, 
but also the important difficulties and contradictions. We now have a much 
clearer idea of the work that needs to be done. In particular the 
theoreticians are going to find in the proceedings of this conference the 
experimental results sufficiently to the point to give their work a 
solid foundation.  \\

I have also been struck by the importance of the new contributions from groups
coming from all parts of the world. Certain physicists, working in the shadows
with ardor and patience, have surprised us with a magnificent burst of results:
I think of Peters, who quietly working in Bombay in Bahbha's Institute very
far from the noise of our western conferences arrived with an extraordinary 
quantity of results that he explained to us methodically during the 40 minutes
allowed for his communication. Then he told us, there are now all our neutral
particles about which I have yet to speak.\\

If we want to draw certain lessons from this congress lets point out first
that in the future we must use the particle accelerators. Lets point out for
example the possibility that they will permit the measurement of certain
fundamental curves (scattering, ionization, range) which will permit us to 
differentiate effects such as the existence of pi mesons among the secondaries
of K mesons.\\

Finally I would like to finish with some words on a subject that is dear to
my heart and is equally so to all the ``cosmicians", in particular the
``old timers". Gradually one becomes an ``old timer". One does not perceive the 
exact moment, but one day one notices that the hair is turning grey and that
the youngsters who were your students are catching up with you and in some 
cases passing you. I can see in the audience some of the "old timers":
The professors Regener, Powell, Blackett,  Vallarta, Bothe, Fretter, ...
We have to face the grave question: what is the future of cosmic rays?
Should we continue to struggle for a few new results or would it be better
to turn to the machines?.\\

 One can no doubt say that that the future of
cosmic radiation in the domain of nuclear physics depends on the machines
especially with the development more or less quickly of ``strong focusing".
But probably this point of view should be tempered by the fact that we
have the uniqueness of some phenomena, quite rare it is true, for which
the energies are much larger (than present machines) and we have a bit of 
protection in that our techniques will be well developed to keep ahead
of the rapid growth of the energy of the machines. We are - I believe -
a bit in the position of a group of alpinists who are climbing a mountain,
this mountain is very high, perhaps infinitely high, and we are climbing it
under conditions more and more difficult. But we cannot stop to take a rest,
to sleep because coming from below a sea is surging, an inundation, a flood
which progressively grows, forcing us to climb higher and higher. It is
evident that this is a position which is not very comfortable, but is it not 
a situation extremely exciting and of magnificent interest?   }\\

\section*{ Views of the conference 30 years later}

Some 30 years later in 1982 a colloquium \cite{1982Colloque} was held on the history of particle physics. The historic role of the
 Bagn\`{e}res de Bigorre was central
in the reminiscences of  R. H. Dalitz and Charles Peyrou \\

 Peyrou writes:\\

{\em The Bagn\`{e}res meeting was the regular cosmic ray conference held every odd year. The organizing committee decided that the main weight of the conference should be on new particles( i. e. strange in modern language). In fact on six full days of conference only one morning was devoted to other topics and the cosmic aspect of cosmic rays was almost completely neglected.The last one and a half days were entirely devoted to a resum\'{e} of the situation coming out of the preceding days. All participants, young or old, remember the conference as the best of their lives. Indeed, it is there that a coherent picture of the new particle physics began to emerge from many partial works. The existence of many particles with well defined properties was firmly and definitely established here, the $\Lambda^0$, the $\theta^0$ (K$^0_1$), the $\Xi^-$. Unsolved problems were at least clearly stated: The nature of the K$^+$ decays, the absence of K$^-$ captures etc. but I should not anticipate too much. In short it is very much due to this conference (and to the first Rochester conferences) that the physics of new(strange) particles began to be considered seriously by "serious" physicists (Serious having, as usual, the sense of either theorists or people busy with classical antiquities and especially physicists combining the two qualities c.f. Gell-Mann's talk at this colloquium).\\

Two years later the cosmos took its revenge. The cosmic ray conference in Mexico city was entirely devoted entirely to topics other than new particles and it is at another conference, at Pisa, that the cosmic rays celebrated a final triumph in that field, only to abandon it definitely to accelerators.}\\

At the same 1982 colloquium Dalitz writes:\\

{\em The International Cosmic ray Conference for 1953 was arranged to concentrate on the new particles, and this was a major event in the lives of all the physicists that took part in it.   It was held 6-12 July at Bagn\`{e}res de Bigorre in the Basque country on the northern slopes of the Pyrenees.  During this conference, it became clear that there was a substantial consensus concerning the subject matter of all this widespread cosmic ray work.  Previously it had seemed that if a new decay mode, or perhaps a known decay mode for a new parent mass, was being reported every month, but now it was seen that the most frequent decay modes were quite limited in number and were associated with fairly definite  mass values. Previously, the V$^0_1$ $\rightarrow$ p+ $\pi^-$ was the only well established
V$^0$ particle, known to us as the $\Lambda$(1115) hyperon; now the painstakingly precise work of Thompson established the
existence of the V$^0_2$ $\rightarrow$ $\pi^+$ $\pi^-$ particle, also known as the $\theta^0$ meson , with mass 496 $\pm$ 5 MeV,
comparable with that for the $\tau^+$ meson. Of course, the cloud chamber work still left some further neutral events, labeled V$^0_3$,
V$^0_4$, ...., and left some charged V$^{\pm}$ events to be sorted out, the latter still to be related with the at-rest decay events reported from the emulsion work.  The new techniques for using layered emulsion blocks were being perfected and it was becoming possible to follow charged secondaries through many layers and so to identify them uniquely and to measure their energies from their range. In committee, rules were drawn up for the formal specification of any new particles or new decay modes. It was an exciting time, as if the mists were lifting and we could at last look ahead.}\\

The contribution of Leprince-Ringuet to the 1982 colloquium is most interesting. He writes: \\

{\em The congress of Bagn\`{e}res de
Bigorre sounded the death knell for  cosmic rays and it was Powell himself who in his closing discourse had said `` Gentleman,
now we are invaded, we are submerged, these are the accelerators'' }\\

In fact Cecil Powell according to the proceedings made no closing statments. The quotation attributed to Powell was in fact
made by Leprince-Ringuet!  Perhaps in 29 years he had forgotten. He continues:\\

{\em Effectively, most of the cosmic ray laboratories including ours at Ecole Polytechnique and College de France oriented 
themselves towards the large particle accelerators and I would like to tell you also that the word hyperon was announced for the first time at the Bagn\`{e}res congress. B. Rossi, E. Amaldi, and C. Powell were there. We asked ourselves how should we call these
new particles which were coming to rest, which were heavy and which were decaying to a meson. Diverse names were
proposed. And I must say that it was my principal contribution to physics that I suggested the word hyperon. The word hyperon was not  well received by Rossi.  Rossi had said ``Oh hyperon, piperon, that wont do''. And on the other hand Powell was there and said
`` Oh hyperon (pronounced haiperon) marvelous'' . And the word hyperon was adopted. And now there is in Bagn\`{e}res de
Bigorre an avenue de l'Hyperon ; it is perhaps the only spot in the world where an avenue has been named for a fundamental particle}.\\

Shown in Figure 5 is a photo of a small card, found in the archives, signed by the group that coined the word hyperon. Note that despite his alleged complaints
Rossi did sign the card.

\begin{figure}[!ht]
\begin{center}
\includegraphics  [width=1.0\textwidth] {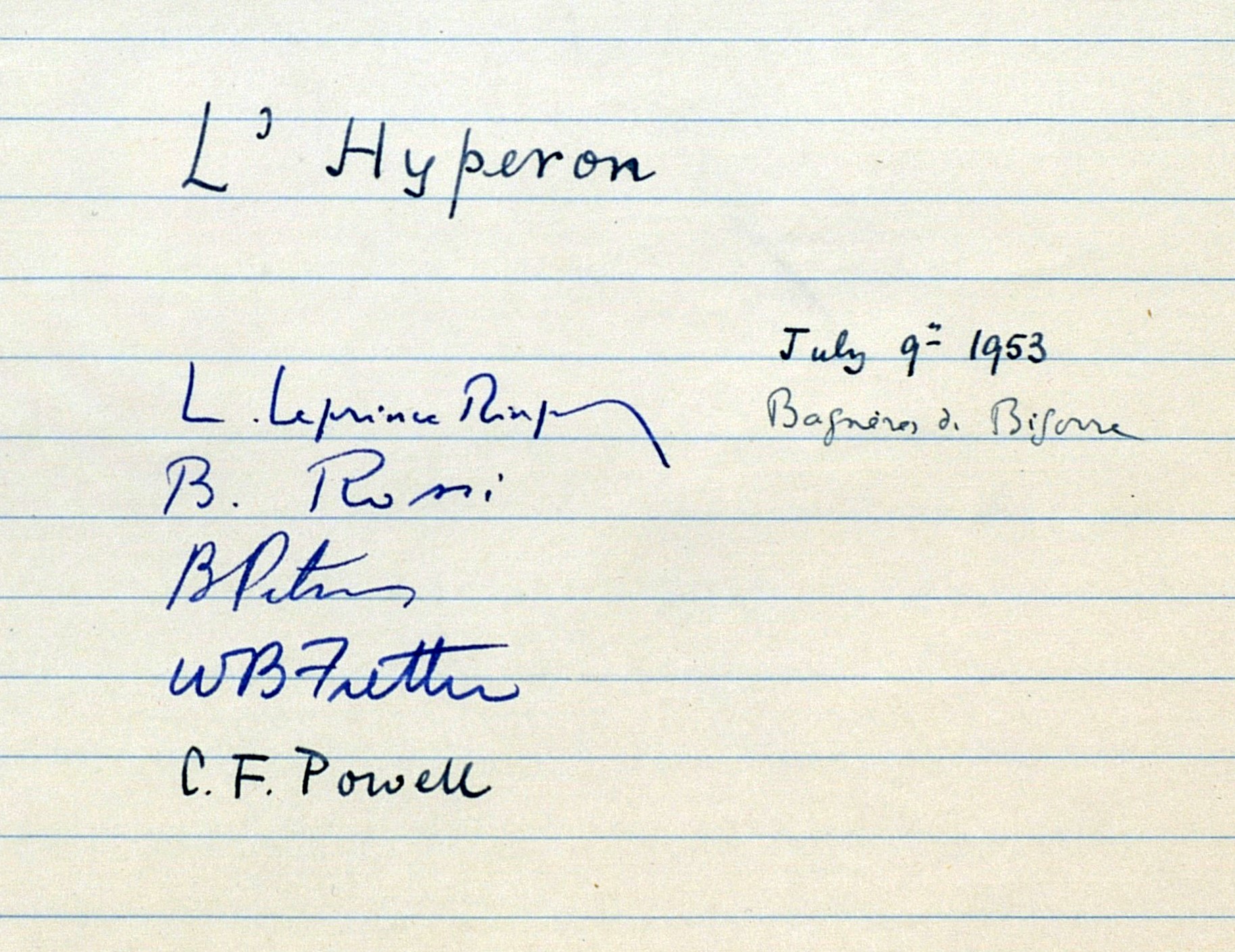}
\end{center}
\caption{\it Card marking the coining the name Hyperon}
\label{fig:Bagn5}
\end{figure}

Leprince-Ringuet seems to make modest fun of his role in the organization of the Bagn\`{e}res congress. In fact it was
his insistence in the narrow focus of the conference that made it a success. The words of of Peyrou and Dalitz provide testimony to that fact.
And Leprince-Ringuet followed his own advice and turned his group to the accelerators which were being built at CERN.  In the 50's
the laboratory directed by Leprince-Ringuet was filled with extraordinary students, Peyrou, Gregory, Lagarrigue, Armenteros, 
 Muller and Astier who became prominent researchers at CERN.

\section*{Concluding Remarks}

 Leprince-Ringuet's conclusion that
the accelerators were about to take over the ``domain of nuclear physics" in cosmic ray research was the correct one. The
Bagn\`{e}res de Bigorre conference marked almost exactly the time of take over by the accelerators. First with the US accelerators Cosmotron and Bevatron followed by the CERN PS and the Brookhaven AGS.
But it was true that the 
general conclusions found in the cosmic radiation concerning the heavy unstable particles were essentially correct. The experiments for the accelerators were already well
defined. In addition the work of Pais, Gell-Mann, Dalitz, and Nishijima done in 1952-53 was correct concerning the systematics of
production and decay of the heavy unstable particles.\\

 Sadly Pais decided at the last minute not to attend so 
there was no presentation by him and no  participation in the discussions. A participation by Gell-Mann at the conference certainly
would have answered some of the lingering questions raised. How well established were the charged hyperons? Why were there
many less negative K mesons than positive ones? Was it really possible that a V$^0_1$ could be produced by a 227 MeV 
$\pi ^-$ ?\\

The success of the conference was due to the extraordinary efforts of Leprince-Ringuet. But it would have even been more successful
if more attention were given to the invitations to the young theoreticians who had  already been thinking about the heavy
unstable particles.  As mentioned above, Leprince invited all the prestigious physicists of the day; however it was a new generation
of physicists who were to understand the cosmic ray discoveries. \\

\section*{Bibliographic note}

All the letters referenced in the text by correspondents and date were found in the Leprince-Ringuet archives at Ecole Polytechnique, Palaiseau, France. They were found in folders contained in boxes 83and 84. The archivist responsible for the organization of the 
Leprince-Ringuet papers was Olivier Azzola. \\

\section* {Acknowledgements}
In preparing this article I had a interviews with individuals who attended the conference at Bagn\`{e}res de Bigorre.
These included Michael Friedlander, Washington University,  Milla Baldo-Ceolin, University of Bologna, Donald Perkins, 
Oxford University and Riccarco Levi-Setti, University of Chicago They all had vivid memories of
the conference. I would also like to thank my colleague Alan Watson of Leeds University for reading the manuscript and making
some suggestions. I thank Wolf Beiglboeck for making some excellent comments and pointing out a factual error
in the early stage of the manuscript.  I thank the archivist of the Leprince-Ringuet papers, Olivier Azzola
for all his assistance on my many trips to Palaiseau. He also performed a great service in placing the rare proceedings
of the Bagn\`{e}res de Bigorre conference on the internet. This research was partially supported by a grant from the France-Chicago Center in the FACCTs program.

\section*{Appendix - The Gell-Mann strangeness scheme }

By 1953 Gell-Mann proposed a scheme for the placement of the newly discovered
heavy unstable particles among the known particles. Most of these particles were referred to in Rossi's summary. 
The scheme is shown in Figure 6 and
includes a new quantum number, strangeness (S).  In the strong (production) interactions S is conserved.
For the weak decays S is not conserved. The new quantum number required associated production.
$\pi^-$ + p  $\rightarrow$  $\Lambda^0$ + K$^0$ is allowed but $\pi^-$ + p  $\rightarrow$  $\Lambda^0$ + $\pi^0$
is not.  K$^-$ mesons can only be produced in pairs with K$^+$ mesons to conserve the strangeness. However
 K$^+$  mesons can be produced in other reactions as well, so Sard's prediction of unequal production was correct. The Gell-Mann scheme accounted
for all the observations of the heavy unstable particles found in the cosmic rays.

\begin{figure}[!ht]
\begin{center}
\includegraphics  [width=0.5\textwidth] {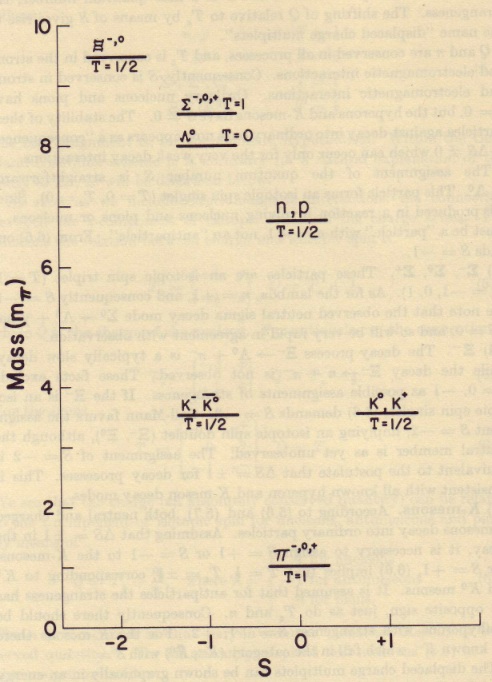}
\end{center}
\caption{\it Gell-mann's organization of the particles with a new quantum number S}
\label{fig:Bagn6}
\end{figure}


\begin{thebibliography}{99}
\bibitem{Schein} The Marcel Schein papers are held in the University of Chicago archives. A thorough search found no
reference to the experimental work supporting Schein's claim that $\Lambda$ hyperons can be produced by low energy negative pions.
\end{thebibliography}
\end{document}